\documentstyle[12pt]{article}
\begin{document}
\thispagestyle{empty}
\newcommand{\be}{\begin{equation}}
\newcommand{\ee}{\end{equation}}
\newcommand{\sect}[1]{\setcounter{equation}{0}\section{#1}}
\newcommand{\vs}[1]{\rule[- #1 mm]{0mm}{#1 mm}}
\newcommand{\hs}[1]{\hspace{#1mm}}
\newcommand{\mb}[1]{\hs{5}\mbox{#1}\hs{5}}
\newcommand{\bea}{\begin{eqnarray}}
\newcommand{\eea}{\end{eqnarray}}
\newcommand{\wt}[1]{\widetilde{#1}}
\newcommand{\ux}[1]{\underline{#1}}
\newcommand{\ov}[1]{\overline{#1}}
\newcommand{\sm}[2]{\frac{\mbox{\footnotesize #1}\vs{-2}}
           {\vs{-2}\mbox{\footnotesize #2}}}
\newcommand{\prt}{\partial}
\newcommand{\eps}{\epsilon}\newcommand{\p}[1]{(\ref{#1})}
\newcommand{\R}{\mbox{\rule{0.2mm}{2.8mm}\hspace{-1.5mm} R}}
\newcommand{\Z}{Z\hspace{-2mm}Z}
\newcommand{\cd}{{\cal D}}
\newcommand{\cg}{{\cal G}}
\newcommand{\ck}{{\cal K}}
\newcommand{\cw}{{\cal W}}
\newcommand{\vj}{\vec{J}}
\newcommand{\vl}{\vec{\lambda}}
\newcommand{\vz}{\vec{\sigma}}
\newcommand{\vt}{\vec{\tau}}
\newcommand{\poiss}{\stackrel{\otimes}{,}}
\newcommand{\tx}{\theta_{12}}
\newcommand{\tb}{\overline{\theta}_{12}}
\newcommand{\zw}{{1\over z_{12}}}
\newcommand{\sqp}{{(1 + i\sqrt{3})\over 2}}
\newcommand{\sqm}{{(1 - i\sqrt{3})\over 2}}
% REVUES POUR BIBLIO
\newcommand{\NP}[1]{Nucl.\ Phys.\ {\bf #1}}
\newcommand{\PLB}[1]{Phys.\ Lett.\ {B \bf #1}}
\newcommand{\PLA}[1]{Phys.\ Lett.\ {A \bf #1}}
\newcommand{\NC}[1]{Nuovo Cimento {\bf #1}}
\newcommand{\CMP}[1]{Commun.\ Math.\ Phys.\ {\bf #1}}
\newcommand{\PR}[1]{Phys.\ Rev.\ {\bf #1}}
\newcommand{\PRL}[1]{Phys.\ Rev.\ Lett.\ {\bf #1}}
\newcommand{\MPL}[1]{Mod.\ Phys.\ Lett.\ {\bf #1}}
\newcommand{\BLMS}[1]{Bull.\ London Math.\ Soc.\ {\bf #1}}
\newcommand{\IJMP}[1]{Int.\ J.\ Mod.\ Phys.\ {\bf #1}}
\newcommand{\JMP}[1]{Jour.\ Math.\ Phys.\ {\bf #1}}
\newcommand{\LMP}[1]{Lett.\ Math.\ Phys.\ {\bf #1}}
%\renewcommand{\thefootnote}{\fnsymbol{footnote}}
%\footnotemark
\newpage
\setcounter{page}{0} \thispagestyle{empty} \vs{12}

\renewcommand{\thefootnote}{\fnsymbol{footnote}}
% {{{\large {\em Draft version}}}
{\par  CBPF-NF-002/02}

\vspace{.5cm}
\begin{center}
{\large\bf Residual Symmetries in the Presence of an EM
Background} \vspace{.5cm}\\

\vs{10} {\large H.L. Carrion, M. Rojas and F. Toppan} ~\\ \quad
\\
 {\large{\em CBPF - CCP}}\\{\em Rua Dr. Xavier Sigaud
150, cep 22290-180 Rio de Janeiro (RJ)}\\{\em Brazil}\\

\end{center}
{\quad}\\ \centerline{ {\bf Abstract}}

\vs{6}

The symmetry algebra of  a QFT in the presence of an external EM
background (named ``residual symmetry") is investigated within a
Lie-algebraic, model independent scheme. Some results previously
encountered in the literature are here extended. In particular we
compute the symmetry algebra for a constant EM background in D=3
and D=4 dimensions. In D= 3 dimensions the residual symmetry
algebra is isomorphic to $u(1)\oplus {\cal P}_c(2)$, with ${\cal
P}_c(2)$  the centrally extended 2-dimensional Poincar\'e algebra.
In D=4 dimension the generic residual symmetry algebra is given by
a seven-dimensional solvable Lie algebra which is explicitly
computed. Residual symmetry algebras are also computed for
specific non-constant EM backgrounds.

\vfill \setcounter{page}0
\renewcommand{\thefootnote}{\arabic{footnote}}
\setcounter{footnote}0
\newpage
\section{Introduction}

The issues of QFTs in an external (constant) EM background gained
interest recently with the Seiberg and Witten's observation
\cite{SW} that an ordinary theory in a constant EM background can
be reformulated as a non-commutative gauge theory.\par The problem
of determining the symmetry algebra of a QFT in a constant EM
background has been addressed and solved, for the very specific
two-dimensional free massive complex boson model minimally coupled
to an external gauge field, in \cite{Kar}. It was proven in that
work that its symmetry algebra coincides with the centrally
extended Poincar\'e algebra in $1+1$ dimension, previously
investigated in a series of papers \cite{{CJ1},{CJ2}}. \par In the
present work we extend the results of \cite{Kar}. By using solely
Lie-algebraic and model-independent methods we compute the
symmetry algebra of different classes of QFTs coupled to a given
external EM background. Throughout the text we call such symmetry
algebras ``residual symmetries". It is worth stressing that, due
to the presence of central extensions, the residual symmetries
{\em are not} subalgebras of the original symmetry algebra in the
absence of the external EM background (such an algebra is given by
the direct sum of the Poincar\'e algebra and a global $U(1)$
charge).\par More specifically, we prove that the residual
symmetry algebra of a three-dimensional Poincar\'e invariant QFT
in a constant EM background is given by the $5$-dimensional
solvable Lie algebra $u(1)\oplus {\cal P}_c(2)$, where ${\cal
P}_c(2)$ is the two-dimensional centrally extended Poincar\'e
algebra whose signature, Euclidean or Minkowskian, is determined
by the relative strength of the constant external electric versus
magnetic field.\par The results of \cite{Kar} in $1+1$ dimensions
are consistently recovered from our own results after performing a
dimensional reduction.
\par Furthermore, we compute the residual symmetry algebra for a
four-dimensional Poincar\'e invariant QFT in a generic constant EM
background. The resulting symmetry is a $7$-dimensional solvable
Lie algebra explicitly presented in formulas (\ref{7dim}).\par The
residual symmetry algebra in the presence of non-constant EM
backgrounds has also been computed in various cases and the
results are here presented.\par The scheme of the paper is the
following. In the next section we illustrate the Lie-algebraic
method which allows, in a model-independent manner, to determine
the residual symmetry generators and the corresponding algebra. In
section ${\bf 3}$ we present the resulting residual symmetry
algebra for a $D=3$ QFT in the presence of a constant EM field. In
section ${\bf 4}$ the residual symmetry algebra is computed for a
QFT in the ordinary $D=4$ Minkowski space-time in the presence of
a constant EM background. In section ${\bf 5}$ the case of a
non-constant EM background is treated in some specific examples.
Finally, in the Conclusions, we make some comments about our work,
drawing attention to its possible applications and outlining the
future investigations.

\section{Residual symmetries and their generators.}

Let us discuss in detail for the sake of simplicity the case of
the residual symmetry for generic Poincar\'e-invariant field
theories in $(2+1)$-dimension, coupled with an external constant
EM background. The generalization of this procedure to
higher-dimensional theories and non-constant EM backgrounds, such
as those studied in Sections ${\bf 4}$ and ${\bf 5}$,  is
straightforward and immediate.\par In the absence of the external
electric and magnetic field, the action ${\cal S}$ is assumed to
be invariant under a $7$-parameter symmetry, given by the six
generators of the $(2+1)$-Poincar\'e symmetry which, when acting
on scalar fields (the following discussion however is valid no
matter which is the spin of the fields) are represented by
\begin{eqnarray}
P_\mu &=& -i\partial_\mu, \nonumber\\ M_{\mu\nu} &=& i(x_\mu
\partial_\nu -x_\nu\partial_\mu),\label{poinc}
\end{eqnarray}
(the metric is chosen to be $+--$), plus a remaining symmetry
generator corresponding to the internal global $U(1)$ charge that
will be denoted as $Z$.
\par
It is further assumed that in the action ${\cal S}$ the dependence
on the classical background field is expressed in terms of the
covariant gauge-derivatives \begin{eqnarray} { D}_\mu &=&
\partial_\mu - ieA_\mu, \nonumber \end{eqnarray} with $e$ the
electric charge.\par
\par
In the presence of constant external electric and magnetic fields,
the $F^{\mu\nu}=\partial^\mu A^\nu-\partial^\nu A^\mu$
field-strength is constrained to satisfy
\begin{eqnarray}
F^{0i} = E^i, &\quad F^{ij} = \epsilon^{ij} B,\label{const}
\end{eqnarray}
where $\mu,\nu =0,1,2$ and $i,j=1,2$.  The fields $E^i$ and $B$
are constant. Without loss of generality the $x^1$, $x^2$ spatial
axis can be rotated so that $E^1\equiv E$, $E^2=0$. Throughout the
text this convention is respected.\par In order to recover
(\ref{const}), the gauge field $A_\mu$ must depend at most
linearly on the coordinates $x^0\equiv t$, $x^1\equiv x$ and
$x^2\equiv y$.\par The gauge-transformation
\begin{eqnarray}
&&A_\mu \mapsto {A_\mu}' = A_\mu +\frac{1}{e}\partial_\mu
\alpha(x^\nu) \label{gtrans}
\end{eqnarray}
allows to conveniently choose for $A_\mu$ the gauge-fixing
\begin{eqnarray}
A_0 &=& 0,\nonumber\\ A_i &=& E_i t -\frac{B}{2}\epsilon_{ij} x^j.
\label{gfix}
\end{eqnarray}
The above choice is a good gauge-fixing since it completely fixes
the gauge (no gauge-freedom is left). It will be soon evident that
the residual symmetry is a truly physical symmetry, independent of
the chosen gauge-fixing.\par Due to (\ref{gfix}), the action
${\cal S}$ explicitly depends on the $x^\mu$ coordinates entering
$A_\mu$. The simplest way to compute the symmetry property of an
action such as ${\cal S}$ which explicitly depends on the
coordinates consists in performing the following trick. At first
$A_\mu$ is regarded on the same foot as the other fields entering
${\cal S}$ and assumed to transform as standard vector field under
the global Poincar\'e transformations, namely
\begin{eqnarray}
{A_\mu}' ({x^\rho}') &=& {\Lambda_\mu}^\nu A_\nu (x^\rho)
\end{eqnarray}
for $ {x^\mu}' = {\Lambda^\mu}_\nu x^\nu + a^\mu $.\par For a
generic infinitesimal Poincar\'e transformation, however, the
transformed $A_\mu$ gauge-field no longer respects the
gauge-fixing condition (\ref{gfix}). In the active transformation
viewpoint only fields are entitled to transform, not the
space-time coordinates themselves. $A_\mu$ plays the role of a
fictitious field, inserted to take into account the dependence of
the action ${\cal S}$ on the space-time coordinates caused by the
non-trivial background. Therefore, the overall infinitesimal
transformation $\delta A_\mu$ should be vanishing. This result can
be reached if an infinitesimal gauge transformation (\ref{gtrans})
$\delta_g(A_\mu)$ can be found in order to compensate for the
infinitesimal Poincar\'e transformation $\delta_P(A_\mu)$, i.e. if
the following condition is satisfied
\begin{eqnarray}
&&\delta(A_\mu) =\delta_P(A_\mu) +\delta_{g}(A_\mu)
=0.\label{delta}
\end{eqnarray}
Only those Poincar\'e generators which admit a compensating
gauge-transformation satisfying (\ref{delta}) provide a symmetry
of the ${\cal S}$ action (and therefore enter the residual
symmetry algebra). This is a plain consequence of the original
assumption of the Poincar\'e and manifest gauge invariance for the
action ${\cal S}$ coupled to the gauge-field $A_\mu$.\par Notice
that the original Poincar\'e generators are deformed by the
presence of extra-terms associated to the compensating gauge
transformation. Let $p$ denote a generator of (\ref{poinc}) which
``survives" as a symmetry in the presence of the external
background. The effective generator of the residual symmetry is
\begin{eqnarray}
{\hat p} &=& p+(\ldots),\nonumber \end{eqnarray} where $(\ldots )$
denotes the extra terms arising from the compensating gauge
transformation associated to $p$. Such $(\ldots )$ extra terms are
gauge-fixing dependent. The ``residual symmetry generator" ${\hat
p}$ can only be expressed in a gauge-dependent manner. However,
two gauge-fixing choices are related by a gauge transformation
${\bf g}$. The residual symmetry generator in the new
gauge-fixing, denoted as ${\tilde p}$, is related to the previous
one by an Adjoint transformation
\begin{eqnarray}
{\tilde p}&=& {\bf g} {\hat p} {\bf g}^{-1}. \end{eqnarray}
Therefore the residual symmetry algebra does not dependent on the
choice of the gauge fixing and is a truly physical
characterization of the action ${\cal S}$.\par The extra-terms
$(\ldots )$ are necessarily linear in the space-time coordinates
when associated with a translation generator, and bilinear when
associated to a surviving Lorentz generator (for a constant EM
background). Their presence implies the arising of the central
term in the commutator of the deformed translation generators.

\section{The residual symmetry for the $(2+1)$ Poincar\'e case.}

The residual symmetry algebra of the $(2+1)$-Poincar\'e theory
involves, besides the global $U(1)$ generator $Z$, the three
deformed translations and just one deformed Lorentz generator (the
remaining two Lorentz generators are broken).\par Within the
(\ref{gfix}) gauge-fixing choice the deformed translations are
explicitly given by
\begin{eqnarray}
P_0 &=& -i\partial_t -e E x,\nonumber\\ P_1 &=&
-i\partial_x-\frac{e}{2}B y,\nonumber\\ P_2 &=& -i\partial_y
+\frac{e}{2}B x.\label{trasl}
\end{eqnarray}
The deformed generator of the residual Lorentz symmetry is
explicitly given, in the same gauge-fixing and for $E\neq 0$, by
\begin{eqnarray}
M &=& i(x\partial_t + t\partial_x) - i\frac{B}{E}
(y\partial_x-x\partial_y) +\nonumber\\&&\frac{e}{2}(E
t^2+Ex^2-Bty).\label{Lor}
\end{eqnarray}
The residual symmetry algebra is given by
\begin{eqnarray}
\relax [ P_0, P_1 ] &=& i E Z,\nonumber\\ \relax [P_0, P_2] &=&
0,\nonumber\\ \relax [P_1,P_2] &=&  i B Z, \nonumber\\
 \relax [M,
P_0]&=& -i P_1,\nonumber\\ \relax [M, P_1] &=& -i P_0 -i
\frac{B}{E} P_2,\nonumber\\ \relax [M, P_2] &=& i\frac{B}{E}
P_1.\label{rsym}
\end{eqnarray}
The $U(1)$ charge $Z$ is no longer decoupled from the other
symmetry generators. It appears instead in (\ref{rsym}) as a
central charge.\par Please notice that the residual symmetry
algebra in $(1+1)$ dimensions (computed in \cite{Kar} for a
specific model) is recovered from the $P_0, P_1, M, Z$ subalgebra.
It corresponds to the centrally extended $2D$ Poincar\'e algebra
thoroughly studied in \cite{CJ2}.
\par The $5$-generator solvable, non-simple Lie algebra of
residual symmetries admits a convenient presentation. The
generator
\begin{eqnarray} {\tilde Z}\equiv B P_0 + E P_2
\end{eqnarray}
commutes with all the other $\ast$ generators
\begin{eqnarray}
\relax [ {\tilde Z}, \ast ] &=&0,
\end{eqnarray}
so that the residual symmetry algebra is given by a direct sum of
$u(1)$ and a $4$-generator algebra. The latter algebra is
isomorphic to the centrally extended two-dimensional Poincar\'e
algebra. Such an algebra is of Minkowskian or Euclidean type
according to whether $ E> B$ or respectively $E<B$ (the case $E=B$
is degenerate). This point can be intuitively understood due to
the predominance of the electric or magnetic effect (in the
absence of the electric field the theory is manifestly rotational
invariant, so that the Lorentz generator is associated with the
Euclidean symmetry). We have explicitly, for $B>E$, that the
algebra
\begin{eqnarray}
\relax [{\overline M}, S_1]&=& i S_2,\nonumber\\ \relax
[{\overline M}, S_2] &=& -i S_1
\end{eqnarray}
is reproduced by
\begin{eqnarray}
{\overline M} &=&  \frac{E}{\sqrt{B^2-E^2}}M,\nonumber\\ S_1 &=&
P_0 + \frac{B}{E} P_2,\nonumber\\ S_2 &=& \frac{\sqrt{B^2-E^2}}{E}
P_1,
\end{eqnarray}
while for $E>B$ the algebra
\begin{eqnarray}
\relax [{\tilde M}, T_1]&=& i T_2,\nonumber\\ \relax [{\tilde M},
T_2] &=& i T_1,
\end{eqnarray}
is reproduced by
\begin{eqnarray}
{\tilde M}&=&\frac{E}{  \sqrt{E^2-B^2}} M,\nonumber\\ T_1 &=& P_0
+ \frac{B}{E} P_2,\nonumber\\ T_2 &=& -\frac{\sqrt{E^2-B^2}}{E}
P_1.
\end{eqnarray}
In both cases the commutator between the translation generators
$S_1$, $S_2$, and respectively $T_1$, $T_2$, develops the central
term proportional to $Z$ which can be conveniently normalized.\par
The residual symmetry algebra of the $(2+1)$ case for generic
values of $E$ and $B$ (the $E=B$ case is degenerate) is therefore
given by the direct sum \begin{eqnarray} &&u(1)\oplus {\cal P}_c
(2).\end{eqnarray} Besides the two charges $Z$, ${\tilde Z}$ an
extra charge is given by the order two Casimir of the centrally
extended Poincar\'e algebra, see \cite{CJ2} for details.

\section{The residual symmetry in $4$ dimensions.}

In $D=4$ dimensions, for generic values of the constant electric
and magnetic field, a convenient gauge-fixing is provided by
\begin{eqnarray}
   A_{0} &=&0 , \nonumber\\
   A_{1} &=& E_{1}  t ,  \nonumber\\
   A_{2} &=& E_{2}  t + B  z ,\nonumber\\
   A_{3} &=& 0.
   \end{eqnarray}
We notice that, without loss of generality, we assume the constant
external magnetic field ${\overrightarrow{B}}$ parallel to the $x$
axis, while $E_1$, $E_2$ denote the components of the external
electric field, respectively parallel and transverse to
${\overrightarrow{B}}$. In the following we consider the case
$E_1, E_2, B\neq 0$.\par The deformed translation generators are
now
\begin{eqnarray}
   P_{0} &=& -i\partial_{t}- e(E_{1} x + E_{2} y), \nonumber  \\
   P_{1} &=& -i\partial_{x}  , \nonumber\\
   P_{2} &=&  -i\partial_{y},   \nonumber \\
   P_{3} &=&     -i\partial_{z} - e B y. \nonumber\\
   \end{eqnarray}
For what concerns the Lorentz generators, only two of them survive
as symmetry generators in the given external background. They are
given by
 \begin{eqnarray}
 M &=& i\frac{B} {E_{1}}  y  \partial_{t}- iz \partial_{x}+i
   \frac{B} {E_{1}} t \partial_{y}
        +i (\frac{E_{1}^{2}+B^{2}} {E_{1} E_{2}})  z \partial_{y}+ix \partial_{z}
        -i(\frac{E_{1}^{2}+B^{2})}{E_{1} E_{2}}) y \partial_{z} +\nonumber \\
      &&  e B (\frac{- E_{1}^{2}+ E_{2}^{2}-B^{2}}{2 E_{1}E_{2}})
        y^{2}+
               e B (\frac {E_{1}^{2}+B^{2}}{2 E_{1}E_{2}})  z^{2}
               +
                   e \frac{B^{2}} {E_{1}}  t z+ e B x y  +
                   e  \frac{E_{2} B} {2 E_{1}}  t^{2},\nonumber\\
   N &=& ix \partial_{t}+i \frac{E_{2}}{E_{1}}  y \partial_{t}+
   it \partial_{x}+  i\frac{E_{2}}{E_{1}}  t \partial_{y}+
   i \frac{B}{E_{1}}  z \partial_{y}-i\frac{B} E_{1}  y
   \partial_{z}+\nonumber\\&&
    \frac{ e}{2  {E_{1}}} ( E_{1}^{2} + E_{2}^{2})  t^{2}
    +\frac{ e}{2}  {E_{1}}  x^{2}+\frac{ e}{2 {E_{1}}}(
 E^2_{2}-B^{2})y^{2} +\frac{ e}{2  {E_{1}}} B^{2}  z^{2} + e B  \frac{E_{2}}{E_{1}}  t z
 + e  {E_{2}}  x y.
\end{eqnarray}
 It is convenient to normalize the generators according to
 \begin{eqnarray}
   T_{1} &=& (1/\sqrt{E_{2}})  P_{0},\nonumber  \\
   T_{2} &=& (1/\sqrt{E_{2}})  P_{2} ,\nonumber \\
   S_{1} &=& (\sqrt{E_{2}}/E_{1})  P_{1},\nonumber  \\
   S_{2} &=& -(\sqrt{E_{2}}/B)  P_{3}.\nonumber
   \end{eqnarray}
The resulting residual symmetry algebra is a three-graded
non-simple solvable Lie algebra, given by the commutators
 \begin{eqnarray}
   \relax [T_{i}, S_{j} ] &=& i\delta_{ij} Z, \nonumber \\
   \relax [T_{i}, T_{j} ] &=& i\epsilon_{ij} Z, \nonumber \\
   \relax [S_{i}, S_{j} ] &=& 0,\nonumber \\
   \relax [M, T_{1} ] &=& -i\frac{B}{E_{1}}   T_{2},\nonumber \\
   \relax [M, S_{1} ] &=& i\frac{B}{E_{1}}   S_{2},\nonumber \\
   \relax [M, T_{2} ] &=& -i(\frac{E_{1}^{2}+B^{2}}{E_{1} E_{2}})
   \frac{B}{E_{2}}   S_{2} - i\frac{B}{E_{1}}   T_{1},\nonumber  \\
   \relax [M, S_{2} ] &=&
    i(\frac{E_{1}^{2}+B^{2}}{E_{1} B})  T_{2}
                         - i\frac{E_{1}}{B}   S_{1},\nonumber  \\
   \relax [N, T_{1} ] &=& -i\frac{E_{1}}{E_{2}}
   S_{1}-i\frac{E_{2}}{E_{1}}
   T_{2},\nonumber \\
   \relax [N, S_{1} ] &=& -i\frac{E_{2}}{E_{1}}   T_{1} ,\nonumber\\
   \relax [N, T_{2} ] &=& -i\frac{B^{2}}{E_{2} E_{1}}
   S_{2}-i\frac{E_{2}}{E_{1}}
   T_{1}, \nonumber\\
   \relax [N, S_{2} ] &=& i\frac{E_{2}}{E_{1}}   T_{2},\nonumber \\
   \relax [M,N ] &=& 0.
   \label{7dim}
   \end{eqnarray}

Such an algebra admits two independent Casimir operators of order
two, given by
 \begin{eqnarray}
 C_{1} &=&  T_{1} T_{1} + 2 \frac{B^{2}}{E_{2}^{2}} T_{1}
 S_{2} - 2 \frac{E_{1}^{2}}{E_{2}^{2}} S_{1} T_{2} + (-1 +
 \frac{B^{2}}{E_{2}^{2}}+\frac{E^2_{1}}{E^2_{2}}) T_{2} T_{2} +
 \nonumber \\
 & & (\frac{B^{4}}{E_{2}^{4}}+ \frac{B^{2} E_{1}^{2}}{E_{2}^{4}})
 S_{2} S_{2} - 2i \frac{B E_{1}}{E_{2}^{2}}{E_{1}} M Z
  + 2i \frac{E_{1}}{E_{2}} NZ,\nonumber\\
 C_{2} &=&  2 \frac{B^{2}}{E_{1}^{2}} T_{1} S_{2} +   S_{1}
 S_{1} -2 S_{1} T_{2} +  ( 1+
  \frac{B^{2}}{E_{1}^{2}}) T_{2} T_{2} +\nonumber \\
 && ( \frac{B^{2}}{E_{1}^{2}} + \frac{B^{2}}{E_{2}^{2}} +
 \frac{B^{4}}{E_{1}^{2}E_{2}^{2}} ) S_{2} S_{2} -  2
i \frac{B}{E_{1}} M Z .
 \end{eqnarray}

\section{Residual symmetries in the presence of a non-constant EM background.}

In the case of a non-constant EM background, the surviving
symmetry generators are further constrained. Some illustrative
cases are reported below. \\{\em i}) Linear external EM field in
$(1+1)$ dimensions.\par For a field $E$, given by
\begin{eqnarray}
E&=& E_1 x+E_2t,
\end{eqnarray}
a convenient gauge-fixing is
\begin{eqnarray}
A_0&=& 0,\nonumber\\ A_1&=& \frac{E_2}{2}t^2+E_1xt.
\end{eqnarray}
There exists only one symmetry generator left, given by
\begin{eqnarray}
P&=& -i\partial_t+i\frac{E_2}{E_1}\partial_x -\frac{e}{2}E_2x^2.
\end{eqnarray}

{\em ii}) Quadratic external EM field in $1+1$ dimensions.\par For
the external field
 \begin{eqnarray}
  E &=& E_{1} x^{2} + E_{2} t^{2} + E_{12} x t,\nonumber
  \end{eqnarray}
the gauge-fixing is given by
  \begin{eqnarray}
  A_{0} &=&0 , \\
  A_{1} &=& E_{1} x^{2} t + \frac{E_{2}}{3} t^{3} + \frac{E_{12}}{2}
  xt^{2}.\nonumber
 \end{eqnarray}
All spatial symmetries are broken unless the condition
 \begin{eqnarray}
  E_{12} &=& 2 \sqrt{E_{1}E_{2}}\nonumber
 \end{eqnarray}
is satisfied. In this particular case, there exists one symmetry
generator given by
 \begin{eqnarray}
  P &=& -i\partial_{t}+ i\sqrt{\frac{E_{2}}{E_{1}}} \partial_{x}-\frac{e}{3} E_{1}
  x^{3}.
 \end{eqnarray}
{\em iii}) Linear external EM field in $2+1$ dimensions.\par In
the most general temporal-gauge case, two independent deformed
translations survive as symmetry generators (all Lorentz
generators are broken), if the external EM field is constrained to
satisfy
\begin{eqnarray}
 E_{1} &=&  \rho^2 x +  \rho D y + F ,\nonumber\\
 E_{2} &=& \rho D x + D y + G ,\nonumber\\
 B &=& B_{1} t + \rho B_{2} x + B_{2} y + B_{3} .
 \end{eqnarray}
($\rho$ arbitrary).
\par
The gauge-fixing is
 \begin{eqnarray}
  A_{0} &=&0,\nonumber  \\
  A_{1} &=&  \rho^2 D t x + \rho D t y +F t,\\
  A_{2} &=&   \rho\frac{B_{2}}{2} x^{2}+\rho D x t + D t
  y+ B_{2} x y + B_{3} x +G t.
  \end{eqnarray}
The symmetry generators
\begin{eqnarray}
P_{0} &=& -i\partial_{t}- \frac{e}{2} \rho^{2} D x^{2} -
\frac{e}{2} D y^{2} - e \rho D x y - e F x - e G y,\nonumber
\\ P_{1} &=& -i\partial_{x}+i \rho \partial_{y} - \frac{e}{2}
B_{2} y^{2} - e  B_{3} y,
\end{eqnarray}
satisfy the centrally extended algebra
\begin{eqnarray}
\relax[ P_0, P_1]&=& -ie(F-\rho G).
\end{eqnarray}
Finally, let us comment that for a conformal theory in $1+1$
dimension, whose original symmetry is $Vir\oplus Vir$, in the
presence of an external background all the symmetry generators
(apart the Poincar\'e generators for the cases already considered)
are broken.

\section{Conclusions}

In this work we have extended the results of \cite{Kar} in two
directions. We showed the model-independent, Lie-algebraic arising
of the result of \cite{Kar} (originally computed for the free
massive complex boson case in $1+1$ dimension, externally coupled
to a constant EM background) and later we computed the residual
symmetries in the presence of constant EM backgrounds for both the
$D=3$ and the ordinary Minkowskian $D=4$ theories.\par For a
constant EM background the residual symmetry of a $D=3$ theory
corresponds to the algebra $u(1)\oplus {\cal P}_c(2)$, where
${\cal P}_c(2)$ is the centrally extended $2D$ Poincar\'e algebra,
widely investigated, both mathematically and in physical
applications, in a series of papers \cite{{CJ1},{CJ2}}. In
\cite{CJ1} it has been applied, e.g., to the construction of
lineal-gravity theories in $1+1$ dimensions. Due to the presence
of the central term, the adjoint representation of ${\cal P}_c(2)$
is not faithful (see \cite{CJ2}). On the other hand a
$4$-dimensional faithful representation is constructed in the
light of the Kirillov's method (see \cite{Bos} for details). This
method is likely to be extended to compute a faithful
representation for the seven-dimensional solvable Lie algebra
(\ref{7dim}) corresponding to the residual symmetry algebra in a
constant EM background in $4D$.\par The residual symmetry algebras
as those computed above play the same role as the ordinary
Poincar\'e algebras, in the case of QFTs living in a given
constant (classical) EM background.\par It is worth mentioning the
connection of such residual symmetry algebras with the arising of
non-commutative structures, due to the presence of the central
term in the commutators of the (deformed) momenta. A corresponding
dual picture can be given which manifests the non-commutativity at
the level of the space-time coordinates. The connection between
such two dual pictures has been fully explored (for a given
specific toy model), e.g., in \cite{Jac} (see also
\cite{LSZ}).\par One of the seemingly most promising line for
future investigations consists in explicitly linking the role of
the ``deformed Poincar\'e generators", as those computed in
section {\bf 5} in the presence of non-constant EM backgrounds,
with the phenomenon of pair-production, observed for specific
linear EM backgrounds such as light-cone external electric fields
\cite{TTW}. This would aim at a Lie-algebraical characterization
of the pair-production phenomenon.\par Finally, the extension of
the above construction to, let's say, supersymmetric theories,
will provide the deformation of the supersymmetry generators in
the presence of an external EM background.

\end{document}